\documentclass[11pt]{article}
\usepackage{graphicx,amsmath,amsthm,mathrsfs,amssymb}
\usepackage[usenames]{color}
\usepackage{ulem}
\usepackage{authblk}

\newcommand{\add}[1]{{\color[rgb]{1,0,0} #1}}
\newcommand{\ee}{\end{equation}}

\newcommand{\reff}[1]{(\ref{#1})}
\newcommand{\beq}{\begin{equation}}
\newcommand{\eeq}[1]{\label{#1}\end{equation}}
\newcommand{\beqa}{\begin{eqnarray}}
\newcommand{\eea}{\end{eqnarray}}
\newcommand{\eeqa}[1]{\label{#1}\end{eqnarray}}
\newcommand{\beg}{\begin{equation*}}
\newcommand{\eeg}{\end{equation*}}

\newcommand{\bsplit}{\begin{split}}
\newcommand{\esplit}{\end{split}}

\title{Generating Einstein gravity, cosmological constant and Higgs mass from restricted Weyl invariance}

\author[1]{Ariel Edery\thanks{aedery@ubishops.ca}}
\author[2]{Yu Nakayama\thanks{nakayama@theory.caltech.edu}}

\affil[1]{Department of Physics, Bishop's University, 2600 College Street, Sherbrooke, Qu\'{e}bec, Canada, J1M 1Z7.\vspace{1em}}

\affil[2]{Walter Burke Institute for Theoretical Physics, California Institute of Technology, Pasadena, California 91125, USA.}

\begin{document}
\date{}
\maketitle
\begin{abstract}
Recently, it has been pointed out that dimensionless actions in four dimensional curved spacetime possess a symmetry which goes beyond scale invariance but is smaller than full Weyl invariance. This symmetry was dubbed {\it restricted Weyl invariance}. We show that starting with a restricted Weyl invariant action that includes a Higgs sector with no explicit mass, one can generate the Einstein-Hilbert action with cosmological constant and a Higgs mass. The model also contains an extra massless scalar field which couples to the Higgs field (and gravity). If the coupling of this extra scalar field to the Higgs field is negligibly small, this fixes the coefficient of the nonminimal coupling $R \Phi^2$ between the Higgs field and gravity. Besides the Higgs sector, all the other fields of the standard model can be incorporated into the original restricted Weyl invariant action. 

\end{abstract}
\setcounter{page}{1}
\newpage
\section{Introduction}

Since its birth \cite{Weyl}, the Weyl symmetry $g_{\mu\nu} \rightarrow \Omega^2(x) g_{\mu\nu}$ acting on the metric has attracted a lot of attention in theoretical physics. The symmetry, however, is too restrictive, and the strict enforcement on the gravity sector leads to conformal gravity, in which the equation of motion is fourth order and suffers from ghosts (without extra boundary conditions \cite{Maldacena}.) In recent work \cite{YuEdery}, it was shown that full Weyl invariance does not actually represent the symmetry of a generic dimensionless action in four-dimensional curved spacetime. The actual symmetry is {\it restricted Weyl symmetry} where the action is invariant under the following transformation: $g_{\mu\nu}\rightarrow \Omega^2 (x) g_{\mu\nu}$, $\psi\rightarrow \Omega^s \psi$ with the conformal factor $\Omega(x)$ obeying the constraint $g^{\mu\nu}\nabla_{\mu}\nabla_{\nu} \Omega(x)\equiv\Box \Omega(x)=0$ (here $g_{\mu\nu}$ stands for the metric and $\psi$ stands for  a generic field with Weyl(conformal) weight of $s$). The constraint on $\Omega(x)$ does not limit it to a constant so that restricted Weyl symmetry is larger than scale invariance (and smaller than full Weyl invariance.) Examples in the recent literature of restricted Weyl invariant actions include the study of the inflationary behavior of $R^2$ gravity in a conformal framework \cite{Lahanas} and in the framework of supergravity \cite{Kounnas}. 

The Einstein-Hilbert action does not possess full Weyl or restricted Weyl symmetry. However, in this work, we show it can emerge starting from a restricted Weyl invariant action. In fact, not only does the Einstein-Hilbert term emerge but also a cosmological constant and the Higgs mass. In short, we will show that $R^2$ gravity coupled to a Higgs sector with no explicit mass is equivalent to the Einstein-Hilbert action with cosmological constant and massive Higgs field. We therefore obtain the standard model of particle physics\footnote{Other fields of the standard model can be incorporated directly as part of the restricted Weyl invariant action and this is discussed at the end of the article.}coupled to standard gravity.  An extra massless scalar field $ \varphi$ also emerges which couples to gravity and the Higgs field. If the coupling of this extra field to the Higgs field is negligibly small experimentally, this fixes the coefficient of the nonminimal coupling $R \Phi^2$ between the Higgs field and gravity. This is in contrast to coupling the standard model to Einstein gravity directly where the coefficient would remain arbitrary.

\section{Restricted Weyl invariant action and its equivalence}
We begin with the restricted Weyl invariant action that includes a Higgs boson field $\Phi$ with no explicit mass, 
\begin{align}
S_0 = \int d^4x \sqrt{-g} \left( \alpha R^2 - \xi R |\Phi|^2 - (\partial_\mu \bar{\Phi} \partial^\mu \Phi) - \lambda |\Phi|^4 \right)\,.  
 \label{rwia}
\end{align}
We are working in the $(-+++)$ metric convention with positive $R$ in the de-Sitter space-time. The above action contains $R^2$ gravity nonminimally coupled to a Higgs field without any dimensionful parameter. It is invariant under the transformations $g_{\mu\nu}\rightarrow \Omega^2 g_{\mu\nu}$, $\Phi\rightarrow \Phi/\Omega$ with $\Box \Omega=0$ \cite{YuEdery}. The other fields of the standard model can be incorporated into the above action while maintaining restricted Weyl invariance (see discussion at the end of the article).

Although the action \eqref{rwia} is simple, its dynamics is disguised in this form due to the higher derivative nature of the $R^2$ term. To analyze the dynamics of the above action in a more transparent manner, we introduce the auxiliary field $\varphi$ and rewrite the restricted Weyl invariant action \eqref{rwia} into the equivalent form 
\begin{align}
S_1 =\int d^4x \sqrt{-g} \left(-\alpha\left(c_1 \varphi + R + \frac{c_2}{\alpha} |\Phi|^2  \right)^2 + \alpha R^2 - \xi R |\Phi|^2 - (\partial_\mu \bar{\Phi} \partial^\mu \Phi) - \lambda |\Phi|^4 \right) \,. 
\end{align}
The first term (which is squared) does nothing after performing the Gaussian integral over $\varphi$ in the path integral i.e. $\int\mathcal{D}\varphi e^{\add{-} i\alpha\,c_1^2 \int d^4x\sqrt{-g}(\varphi - f(x))^2} = \mathrm{const}$. We take $\varphi$ to be dimensionless. The parameter $c_1$ has therefore dimensions of $\text{(length)}^{-2}$ and $c_2$ is dimensionless. Both $c_1$ and $c_2$ are arbitrary and do not affect any physics.

Expanding  the square in brackets, we obtain 
\begin{align}
S_2 &= \int d^4x \sqrt{-g} \left(-c_1^2\alpha \varphi^2 - 2\alpha c_1 \varphi R  \right. \cr
& \left.  -(\xi+2c_2)R|\Phi|^2 -  (\partial_\mu \bar{\Phi} \partial^\mu \Phi)-2c_1c_2 \varphi |\Phi|^2  -(\alpha^{-1} c_2^2 +\lambda)|\Phi|^4   \right) \ . 
\label{S2}
\end{align}

The above action maintains restricted Weyl invariance if the auxiliary field $\varphi$ transforms as $\varphi\rightarrow \varphi/\Omega^2$ i.e. the action is invariant under the transformations $g_{\mu\nu}\rightarrow \Omega^2 g_{\mu\nu}$, $\Phi\rightarrow \Phi/\Omega$ and $\varphi\rightarrow \varphi/\Omega^2$ with $\Box \Omega=0$. Note that the above action \reff{S2} is restricted Weyl invariant and equivalent to action \reff{rwia} despite the fact that it contains a dimensionful constant $c_1$.

\subsection{Weyl transformation}
We now perform the Weyl transformation
\begin{align}
g_{\mu\nu} &\to \varphi^{-1}g_{\mu\nu} \cr
\sqrt{-g} &\to \varphi^{-2} \sqrt{-g}  \cr
R & \to \varphi R - 6 \varphi^{3/2} \Box \varphi^{-1/2}  \cr
\Phi &\to \varphi^{1/2} \Phi \ .  \label{Weyltransf}
\end{align}

In this new frame, the above action becomes
\begin{align}
S_3 &= \int d^4x \sqrt{-g} \left(-\alpha c_1^2 -2\alpha c_1 \left(R -6\varphi^{1/2} \Box \varphi^{-1/2}\right) \right. \cr
& \left. -(\xi+2c_2)(R-6\varphi^{1/2}\Box\varphi^{-1/2})|\Phi|^2 - (\partial_\mu \bar{\Phi} \partial^\mu \Phi) - \varphi^{1/2} \Box \varphi^{-1/2}|\Phi|^2 \right. \cr
& \left. -2c_1c_2 |\Phi|^2  -(\alpha^{-1} c_2^2 +\lambda)|\Phi|^4    \right) \ .\cr
&=\int d^4x \sqrt{-g} \,\Big(-\alpha c_1^2 -2\,\alpha c_1 R -  \partial_\mu \bar{\Phi} \partial^\mu\Phi -2\,c_1c_2 |\Phi|^2  -(\alpha^{-1} c_2^2 +\lambda)|\Phi|^4 -(\xi+2c_2)R\,|\Phi|^2 \cr& + 3\,\alpha c_1 \dfrac{1}{\varphi^2} \partial_\mu \varphi \,\partial^\mu \varphi +(6(\xi+2c_2)-1)\varphi^{1/2} \Box \varphi^{-1/2}|\Phi|^2\Big)\,. 
\label{S3}
\end{align} 
Now except for the existence of another scalar $\varphi$, the resulting action can be regarded as the Higgs sector of the standard model coupled to Einstein gravity with cosmological constant. The term $-2\,\alpha c_1 R$ is the Einstein-Hilbert term with $-2 \alpha c_1$ determining Newton's constant (in our convention $\alpha c_1$ is negative) The term $-\alpha \,c_1^2$ determines the cosmological constant and we are free to choose its value by setting the value of $c_1$. The physical Higgs mass is determined by $-2 \,c_1\,c_2$ and can be adjusted freely by choosing an appropriate value for $c_2$. The coefficient $-(\alpha^{-1} c_2^2 +\lambda)$ of the quartic term can be adjusted freely by choosing an appropriate value for $\lambda$. Finally, we are free to choose $\xi$ to fix the coefficient of of the nonminimal coupling term $R \Phi^2$. The remaining terms are those involving the auxiliary field $\varphi$. It includes a kinetic term and a coupling to the Higgs field $\Phi$. It is worth mentioning that the cosmological constant does not emerge from the theory of $R+ \alpha R^2$ which had been shown to be equivalent to Einstein gravity plus a massive scalar field (but no cosmological constant) \cite{Whitt}. In contrast, early studies of pure $R^2$ gravity had shown a cosmological constant to emerge \cite{Higgs, Stephenson} in agreement with our work (see also 
\cite{Ferraris} for a general study of pure $R^{n/2}$ gravity where $n$ is the spacetime dimension).    
 
The restricted Weyl symmetry of our original action \reff{rwia} manifests itself in the final action \reff{S3} through a symmetry under a transformation of the auxiliary field $\varphi$ only. The action \reff{S3} is invariant under $\varphi\to \varphi/\Omega^2$, $g_{\mu\nu}\to g_{\mu\nu}$, $\Phi \to \Phi$ with condition $\Box \Omega -\partial_\mu (\ln \varphi) \partial^\mu \Omega =0$. This can be understood as follows. We replaced $g_{\mu\nu}$ with  $\hat{g}_{\mu\nu} = \varphi^{-1} g_{\mu\nu}$. Originally, the restricted Weyl symmetry acts as $\hat{g}_{\mu\nu} \to \hat{g}_{\mu\nu} \Omega^2$, $\varphi \to \varphi/\Omega^2$ and $\Phi\to \Phi/\Omega$ with $\hat{\Box} \Omega =0$. After the replacement, it acts as $g_{\mu\nu} \to g_{\mu\nu}$, $\Phi \to \Phi$, $\varphi \to \varphi/\Omega^2$ with the condition $\hat{\Box} \Omega =0$. In terms of $g_{\mu\nu}$, the condition $\hat{\Box} \Omega =0$ is $\Box \Omega -\partial_\mu (\ln \varphi) \partial^\mu \Omega =0$. In particular note that $\varphi^{1/2} \Box \varphi^{-1/2}$ is invariant under $\varphi \to \varphi/\Omega^2$ when the condition $\Box \Omega -\partial_\mu (\ln \varphi) \partial^\mu \Omega =0$ holds.
This symmetry forbids the mass term for $\varphi$ and gives a severe constraint on how it couples to the other fields. We also realize that when $\varphi$ is non-zero, the original restricted Weyl symmetry is spontaneously broken, and the $\varphi$ transformation in our final action is regarded as the {\it non-linear} realization of the restricted Weyl symmetry. In a certain sense, it is similar to the Dirac-Born-Infeld (DBI) action or Galileon action \cite{Nicolis:2008in}.\footnote{However, we note that our final action is more constrained than just random field theories with extra $\varphi$ field that preserves this symmetry on $\varphi$. The further constraint comes from the fact that a change of the parameter $c_2$ may be compensated by field redefinitions so that it would not affect the physics. See also the discussions on the fermion mass at the end of the article.} 

The auxiliary field $\varphi$ couples to the Higgs field and gravity. If its coupling to the Higgs field is experimentally small, we are free to eliminate it by choosing $\xi+2c_2$ to be $1/6$. The action \reff{S3} then simplifies to
\begin{align}
S_4 &=\int d^4x \sqrt{-g} \,\Big(-\alpha c_1^2 -2\,\alpha c_1 R -  \partial_\mu \bar{\Phi} \partial^\mu\Phi -2\,c_1c_2 |\Phi|^2  -(\alpha^{-1} c_2^2 +\lambda)|\Phi|^4 - \dfrac{1}{6}R\,|\Phi|^2 \cr& + 3\,\alpha c_1 \dfrac{1}{\varphi^2} \partial_\mu \varphi \,\partial^\mu \varphi\Big)\,.
\label{S4}
\end{align} 

Therefore, eliminating the coupling between the auxiliary field $\varphi$ and the Higgs field $\Phi$ {\it fixes} the coefficient for the nonminimal coupling between gravity and the Higgs field. In other words, the coefficient of $R \Phi^2$ is no longer arbitrary but fixed to be ${-}1/6$. This corresponds to conformal coupling. The kinetic term for $\varphi$  can be made canonical by defining the scalar field $\psi \equiv \sqrt{{-}3 \alpha c_1}\,\ln \varphi$ so that $3\,\alpha c_1 \dfrac{1}{\varphi^2} \partial_\mu \varphi \,\partial^\mu \varphi={-}\partial_\mu \psi \,\partial^\mu \psi$. We therefore have a massless minimally coupled scalar field $\psi$ propagating in curved spacetime.  

We now briefly discuss the inclusion of  the other standard model fields into the original action \reff{rwia}. They are collectively given by gauge fields $A_\mu$ with gauge group  $G = U(1)\times SU(2) \times SU(3)$  and (Weyl) fermions $\Psi$ under various representations  of $G$. We may consider the (restricted) Weyl invariant action schematically given by 
\begin{align}
{-}\int d^4x \sqrt{-g}\left( \frac{1}{2g^2}\,\mathrm{Tr} F_{\mu\nu} F^{\mu\nu}  + \theta \,\mathrm{Tr} \epsilon^{\mu\nu\rho\sigma}F_{\mu\nu} F_{\rho\sigma}  +  \bar{\Psi} D_\mu \gamma^\mu \Psi + \lambda(\Psi \Psi \Phi) + \bar{\lambda}(\bar{\Psi}\bar{\Psi}\bar{\Phi}) + \text{Higgs}\ \right) . \label{other}
\end{align}
The Higgs sector is essentially given in \eqref{rwia} by replacing the derivative $\partial_\mu$ with the gauge covariant one.
The gauge coupling constant $g$, the theta angle $\theta$, and Yukawa coupling $\lambda$ are all dimensionless and it is easy to see that the above action does not change under the Weyl transformation \eqref{Weyltransf} supplemented by $\Psi \to \varphi^{3/4}\Psi$. Note that there is no direct coupling to the extra massless scalar field $\varphi$ (except for the Higgs sector we have already discussed). Some beyond the standard model fields such as dark matter and the inflaton may be introduced in a similar way. It is important that the standard model action does not contain Dirac or Majorana type fermion mass terms. If it had such mass terms, we would not be able to rewrite the standard model action into the (restricted) Weyl invariant form such as \eqref{other}.  The neutrino mass term must therefore be introduced in a slightly non-trivial way (e.g. coupling to scalar field with a vaccuum expectation value e.g. in Grand Unified Theories) in order to avoid the direct Dirac or Majorana type mass. 

\section{Conclusion}

In this paper we have shown that starting with a restricted Weyl invariant action that is composed of $R^2$ gravity nonminimally coupled to the Higgs field $\Phi$, one could generate Einstein gravity with cosmological constant and massive Higgs field. Besides the Higgs sector, one could also include into this picture the other fields of the standard model (except for Dirac or Majorana type fermion mass terms which would have to be introduced in a different fashion).     

Our construction accommodates (but not explains) the hierarchy of the scales in the standard model. In contrast to the similar construction in \cite{Kounnas}, the scale of the Higgs field can be hierarchically different from the size of the cosmological constant. This is because our scalar in the original action did not have the conformal kinetic term as theirs. From our philosophy of the restricted Weyl invariance, there is no reason to commit ourselves to the conformal value $\xi = 1/6$.

We would like to conclude by clarifying one point. Action \reff{rwia} was Weyl-rescaled to action \reff{S3}. However, we did not have a mass scale in our original restricted Weyl invariant action \reff{rwia} but we do have one in our final action \reff{S3}. How can we understand this? The answer is that all we observe in experiments are the ratios of dimensionful quantities, the size of the universe in the unit of Planck length (or light-year), the Higgs mass in the unit of Planck mass (or electron-volt) and so on. All these dimensionless parameters are encoded in our original restricted Weyl invariant action. What we found that is non-trivial is that the structure of the dimensionless parameters that we know in the standard model today is consistent with the extra symmetry of restricted Weyl invariance.\footnote{We predict the non-minimal coupling of the Higgs field to gravity as well as an extra massless minimally coupled scalar field, so our model is falsifiable in principle.}
This does not apply to other random field theories.

\section*{Acknowledgments}
A.E. acknowledges support from a discovery grant of the National
Science and Engineering Research Council of Canada (NSERC).  
Y.N. is supported by Sherman Fairchild Senior Research Fellowship at California Institute of Technology and DOE grant number DE-SC0011632. Y.N. thanks Valerio Faraoni for discussion.

\end{document}